\documentclass{article}

\usepackage[final]{neurips_2020}

\usepackage[utf8]{inputenc} % allow utf-8 input
\usepackage[T1]{fontenc}    % use 8-bit T1 fonts
\usepackage{hyperref}       % hyperlinks
\usepackage{url}            % simple URL typesetting
\usepackage{booktabs}       % professional-quality tables
\usepackage{amsfonts}       % blackboard math symbols
\usepackage{nicefrac}       % compact symbols for 1/2, etc.
\usepackage{microtype}      % microtypography

\usepackage{microtype}
\usepackage{graphicx}
\usepackage{wrapfig}
\usepackage{subcaption}
\usepackage{booktabs} % for professional tables

\usepackage{amsmath}
\usepackage{amssymb}
\usepackage{mathrsfs}
\usepackage{amsthm}
\usepackage{stmaryrd}
\usepackage{mathtools}

\usepackage{bm}
\usepackage{dblfloatfix}
\usepackage[noend]{algorithm, algorithmic}

\usepackage{subcaption}
\usepackage{multirow}
\usepackage{xcolor}

\title{GANterpretations}

\author{%
  Pablo Samuel Castro \\
  Google Research, Brain Team\\
  \texttt{psc@google.com} \\
}

\begin{document}

\maketitle

Since the introduction of Generative Adversarial Networks (GANs) \citep{goodfellow14gan} there has been a regular stream of both technical advances (e.g., \cite{arjovsky17wasserstein}) and creative uses of these generative models (e.g., \citep{karras19style,zhu17unpaired,jin2017automatic}). In this work we propose an approach for using the power of GANs to automatically generate videos to accompany audio recordings by aligning to spectral properties of the recording. This allows musicians to explore new forms of multi-modal creative expression, where musical performance can induce an AI-generated musical video that is guided by said performance, as well as a medium for creating a visual narrative to follow a storyline (similar to what was proposed by \cite{frosst19text}).

The ability of GANs to generate a seemingly infinite amount of images is due largely to the fact that training them involves learning a {\em latent space} $Z\in\mathbb{R}^d$, where each point $z\in Z$ generates a unique image $G(z)$, where $G$ is the {\em generator} of the GAN. When trained properly, these latent spaces are learned in a structured manner, where nearby points generate similar images. This enables smooth interpolations between two distinct images (points) $z_1,z_2\in Z$ via their linear combination $G((1 - \alpha) z_1 + \alpha z_2)$ for some $\alpha\in [0, 1]$.
For our work we make use of the BigGAN family of models \citep{brock2018large}, which are class-conditional generative models. In high-level terms, this means that the latent space is of the form $Z\times C$, where $Z\in\mathbb{R}^d$ and $C$ is a finite set of possible categories (such as {\em jellyfish}, {\em bike}, and {\em boa}). These models thus allow us to perform smooth interpolations between images in the same category $c\in C$ ($G((1 - \alpha) (z_1, c) + \alpha (z_2, c))$), and images from different categories $c_1,c_2\in C$ ($G((1 - \alpha) (z_1, c_1) + \alpha (z_2, c_2))$).
Given an audio recording $a$, we compute the spectrogram (the spectrum of different audio frequencies) which produces a 2-dimensional array $S(a)$ of size $F\times T$, where $F$ is the number of frequencies per time sample, and $T$ is the number of time segments the audio file has been decomposed into.

\begin{figure}[!h]
  \centering
  \includegraphics[width=0.7\textwidth]{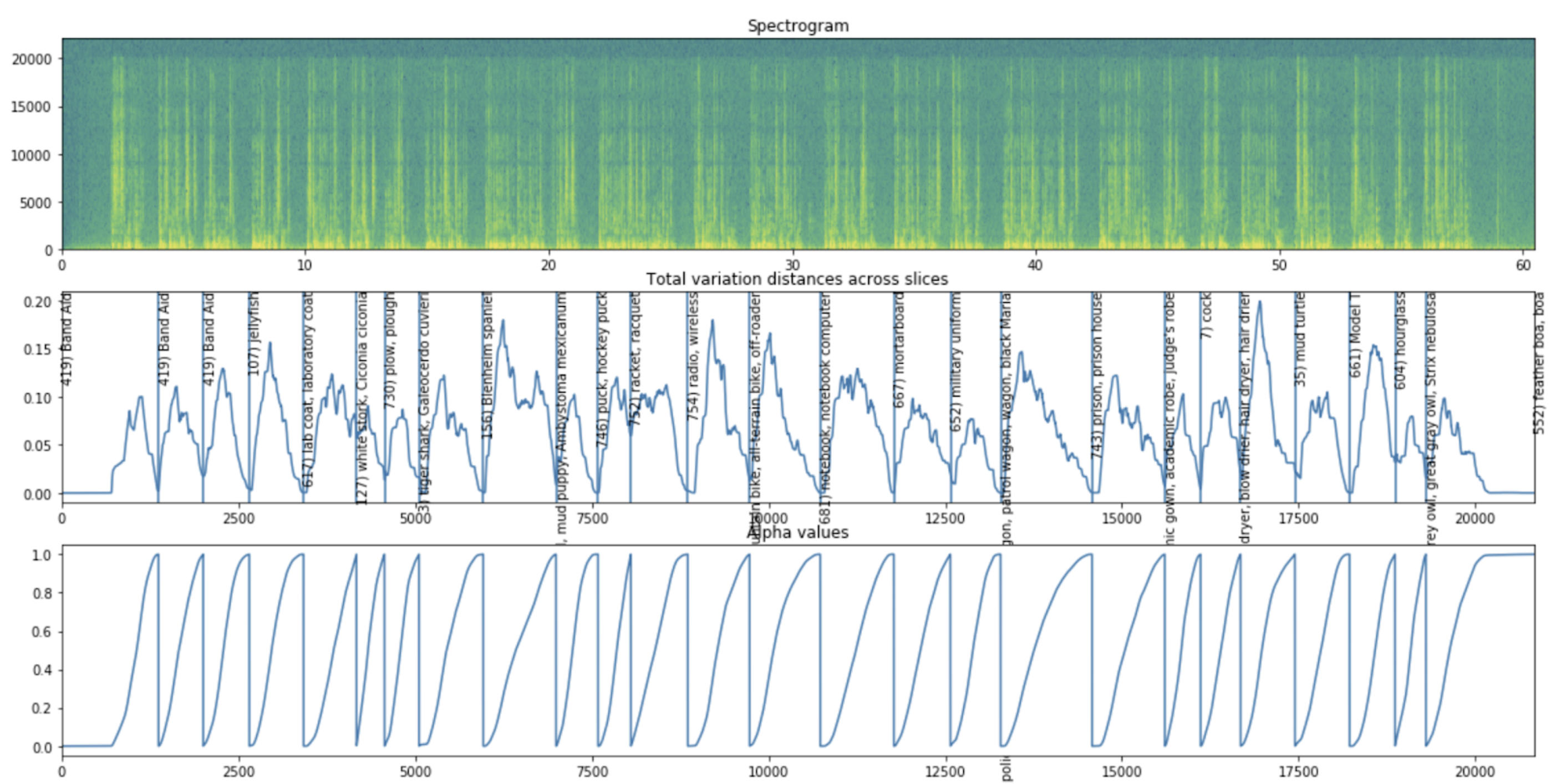}
  \includegraphics[width=0.7\textwidth]{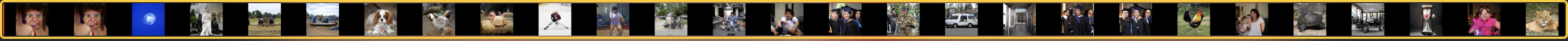}
  \caption{From top to bottom: Audio spectrogram; TV distances with computed inflection points and selected categories; resulting $\alpha$ values used for interpolation; Final video clips}
  \label{fig}
\end{figure}

Our procedure is summarized in Algorithm~\ref{alg:ganterpretation}. In words, it proceeds as follows: {\bf (1)} Compute the Total Variation (TV) distance between the frequency spectrograms of consecutive time slices (line 2). {\bf (2)} Slide through these TV distances and find the inflection points by comparing the current point $t$ with the average TV difference of a window of slice $L$ before and after $t$ (lines 5-6); if the TV distance at point $t$ is at a {\em peak}, then record this as an $inflectionPoint$ (lines 7-9); the first and last points are also added as inflection points (lines 3, 11). {\bf (3)} Compute alpha values by normalizing the cumulative sum between inflection points (lines 13-17). {\bf (4)} For each inflection point, generate a random category or use the ones provided by the user (lines 18 - 24); concurrently, generate latent codes $z$ for each category. {\bf (5)} Using the generated latent points, categories, and interpolations, generate the frames for the video (lines 25-32). This process is illustrated in \autoref{fig}, and the full code is available at {\color{blue} https://github.com/psc-g/ganterpretation}.

We present two demonstrations of this method which allow readers to understand its functionality. In the first, we explore creating a video accompaniement for a musical performance, where the categories were selected randomly ({\color{blue} https://youtu.be/oQI8zG0WNuI}); In the second we demonstrate the ability for visual narrative, where the categories were pre-selected to match the storyline ({\color{blue} https://youtu.be/YelauzLHI6E}).

The method of GANterpretatinos can be used in different ways, two of which are shown in the videos above. Other variations might include varying the method of converting the 2D spectrogram to a 1D signal (instead of via TV distances), using different algorithms for selecting inflection points can be used, and using different GAN models. Additionally, the way the audio is generated/recorded can affect the generated video; indeed, the audio storytelling example used explicit silences to force inflection points at desired times. An avenue we are exploring is pre-generating images in order to have real-time response so as to be used during musical performance; in this way, we will have a closed creative loop: the musical performance affects the video generation, which in turn affects the musical performance.

\begin{algorithm}[H]
\caption{GANterpretations}\label{alg:ganterpretation}
\begin{algorithmic}[1]
  \STATE \textbf{Given}: Audio $a$, rolling length $L$, inflection threshold $\delta$, pre-selected categories $\kappa$ (optional)
  \STATE $TV\leftarrow [mean(|S(a)[:, i] - S[:, i+1]|) \quad {\textrm for } \quad i\in [0, T-1]]$
  \STATE $inflectionPoints\leftarrow [0]$
  \FOR{$t = L+1, \ldots, T-L$}
    \STATE $prevMean \leftarrow mean(TV[t-1-L:t-1]$ (rolling mean before position $t$)
    \STATE $nextMean \leftarrow mean(TV[t+1:t+1+L]$ (rolling mean after position $t$)
    \IF{$sign(TV[t] - prevMean) == sign(TV[t] - nextMean)$ and\\
        $\quad|TV[t] - prevMean| > \delta$ and $|TV[t] - nextMean| > \delta$}
      \STATE Add $t$ to $inflectionPoints$
    \ENDIF
  \ENDFOR
  \STATE Add $T-1$ to $inflectionPoints$
  \STATE $alphas\leftarrow [0.0]$
  \FOR{$i = 1,\ldots, len(inflectionPoints)$}
    \STATE $interpolation \leftarrow cumsum(TV[inflectionPoints[i-1], TV[inflectionPoints[i])$
    \STATE $interpolation /= (inflectionPoints[i] - inflectionPoints[i-1])$
    \STATE Append $interpolation$ to $alphas$.
  \ENDFOR
  \STATE $zs\leftarrow []$
  \FOR{$i = 0,\ldots, len(inflectionPoints)$}
    \STATE Add a random sample from $Z$ to $zs$
    \IF{$\kappa[i]$ does not exist}
      \STATE Add a random category from $C$ to $\kappa$
    \ENDIF
  \ENDFOR
  \STATE $vidFrames\leftarrow []$
  \FOR{$i= 1,\ldots, len(inflectionPoints)$}
    \FOR{$t=inflectionPoints[i-1],\ldots,inflectionPoints[i]$}
      \STATE $\alpha\leftarrow alphas[t]$
      \STATE Add $G((1 - \alpha)(z[i-1], \kappa[i-1]) + \alpha(z[i], \kappa[i]))$ to $vidFrames$
    \ENDFOR
  \ENDFOR
  \STATE Return $vidFrames$
\end{algorithmic}
\end{algorithm}

{\bf Ethical implications}

This work is meant to be for purely artistic use, so the same ethical considerations that apply for all forms of art apply equally to this method. Additionally, as GANs can sometimes overfit to its training data, care must be used if displaying the generated images publicly. For our work we have restricted ourselves to BigGAN, which is trained on the well-known and public ImageNet dataset.

\bibliographystyle{plainnat}
\bibliography{ganterpretations}

\end{document}